\documentclass[preprint,showpacs,preprintnumbers,amsmath,amssymb,showkeys]{revtex4}

\usepackage{graphicx}% Include figure files
\usepackage{dcolumn}% Align table columns on decimal point
\usepackage{bm}% bold math

\begin{document}
%\setlength{\textheight}{7.7truein}  %for 2nd page o
%\def\bea{\begin{eqnarray}}
%\def\eea{\end{eqnarray}}
%\def\nn{\nonumber}
%\draft
%\preprint{}

\title{Ansatz of Leptonic Mixing: The Alliance of Bi-Maximal Mixing with a Single-Angle Rotation.}

%\thispagestyle{empty}
%\begin{titlepage}

\author{Kim Siyeon} \email{siyeon@cau.ac.kr}
\affiliation{Department of Physics,
        Chung-Ang University, Seoul 156-756, Korea}

\date{August 13, 2012}
\begin{abstract}
We introduce an ansatz of the PMNS matrix that consists of specific types of transformations. Bi-maximal mixing is taken for the neutrino masses, while a single-angle rotation in the 1-2 block is taken for the charged lepton masses. Motivated by the implications of the recent results on neutrino oscillations, $\theta_{23}$ in the first octant and non-zero $\theta_{13}$ are predicted by the ansatz. Three physical mixing angles are expressed in terms of a single variable, the 1-2 angle of charged leptons, so that a simple relation among the angles has been obtained: $\tan\theta_{13}=\sqrt{2}(\sin\theta_{23}-\sin\theta_{12})$. It follows that a model of the inverted hierarchy that can produce the given ansatz is proposed.
\end{abstract}


%\end{titlepage}
\maketitle \thispagestyle{empty}

%%%%%%%%%%%%%%%%%%%%%%%%%%%%%%%%%%%%%%%%%%%%%%%%%%%%%%%%%%%%

\section{Introduction}

A series of recent oscillation experiments, T2K\cite{Abe:2011sj}, MINOS\cite{Adamson:2011qu}, Double Chooz\cite{Abe:2011fz}, Daya Bay\cite{An:2012eh}, and RENO\cite{Ahn:2012nd}, made a forward movement in neutrino physics \cite{Beringer:1900zz}. A 3$\nu$ global analysis\cite{Fogli:2012ua} presented the best fit and the allowed ranges of masses and mixing parameters at 90\% confidence level(CL) from the contribution of those experiments. According to the analysis, the $3\sigma$ ranges of the physical parameters are given as in the following: $6.99<\Delta m_{21}^2/10^{-5}eV^2<8.18, ~ 2.59<\sin^2\theta_{12}/10^{-1}<3.59, ~2.19(2.17)<\Delta m_{32}^2/10^{-3}eV^2<2.62(2.61), ~1.69(1.71)<\sin^2\theta_{13}/10^{-2}<3.13(3.15),$ and $3.31(3.35)<\sin^2\theta_{23}/10^{-1}<6.36(6.63)$ for normal(inverted) hierarchy. Besides the non-zero value of $\sin\theta_{13}$, the current data takes the preference for $\theta_{23}$ in the first octant and the inverted mass hierarchy. In the past, due to the central values of $\theta_{23}$ close to $\pi/4$, and due to the absence of the direct measurement of $\theta_{13}$, tri-bi-maximal mixing\cite{Harrison:2002er,Harrison:2003aw} has been the strongest candidate for the Pontecorvo-Maki-Nakagawa-Sakata(PMNS) matrix.

After definitive measurement of $\theta_{13}$ with Daya Bay and RENO, the non-zero $U_{e3}$ has been targeted as a group-theoretical consequence rather than as a perturbed effect. In this paper, an ansatz of PMNS is introduced, which consists of a certain choice of the transformation, $U_l$, of charged lepton mass matrix and bi-maximal mixing matrix, $U_{BM}$, for the neutrino masses. The $U_l$ in the ansatz is a single rotation of the 1-2 block. It will be shown that the three angles in the PMNS matrix are fit into the allowed ranges in the global analysis when the bi-maximal mixing is combined with a single-angle rotation by $U_l^\dagger U_{BM}$. The specific choice of such $U_l$ guarantees the $\theta_{23}$ in the first octant. For comparison, it will also be shown that another ansatz with a single rotation of the 1-3 block always results in the $\theta_{23}$ in the second octant. All three physical mixing angles are given in terms of a single parameter, the angle of the 1-2 rotation, so that it is possible to express the following relation among those angles: $\tan\theta_{13}=\sqrt{2}(\sin\theta_{23}-\sin\theta_{12})$.

A non-Abelian discrete group $\mathbb{S}_3$ is still a good symmetry for the bi-maximal mixing, as it was for the tri-bi-maximal mixing \cite{Meloni:2010aw,Yang:2011fh,Morisi:2011ge,Park:2011zt,Morisi:2011pm,Chu:2011jg,Dong:2011vb,Siyeon:2012zu}. In our model, an Abelian $\mathbb{Z}_2$ symmetry is used to keep the Standard Model(SM) particles separated from the interactions of extra particles beyond the SM. The light mass matrix of neutrinos with bi-maximal mixing and the charged lepton mass with the mixed 1-2 block alone can be built up under $\mathbb{S}_3\otimes\mathbb{Z}_2$ flavor symmetry. The model that gives rise to the PMNS by the product of a single rotation and tri-bi-maximal mixing, $U_l^\dagger U_{TBM}$, is presented in Ref.\cite{Siyeon:2012zu}. There is an article which examines $U_l^\dagger U_{TBM}$ and $U_l^\dagger U_{BM}$ with a single-angle $U_l$ in Ref.\cite{Duarah:2012bd}, while our work has been prepared independently.

This paper is organized as follows: Section II describes the implications of the recent results of neutrino experiments, and introduces the ansatz motivated from those implications. In Section III, a model for the $\theta_{23}$ in the first octant, non-zero $\theta_{13}$ and the inverted mass hierarchy is constructed to support the proposed ansatz. The conclusion section includes a brief summary of the content and the prediction of the ansatz.

\section{Phenomenological conditions and ansatz of PMNS matrix}

\begin{figure}
\resizebox{80mm}{!}{\includegraphics[width=0.75\textwidth]{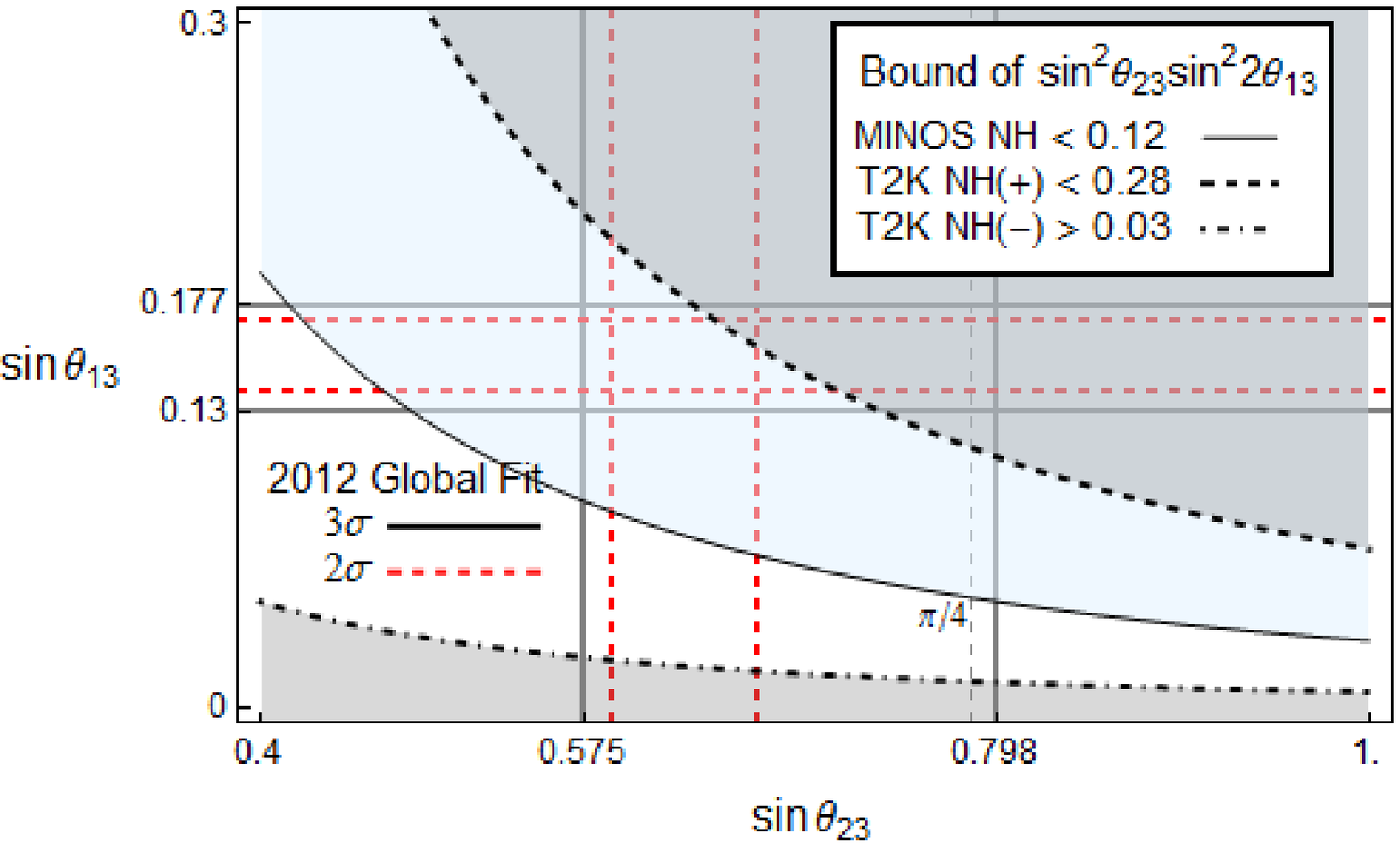}}
\resizebox{80mm}{!}{\includegraphics[width=0.75\textwidth]{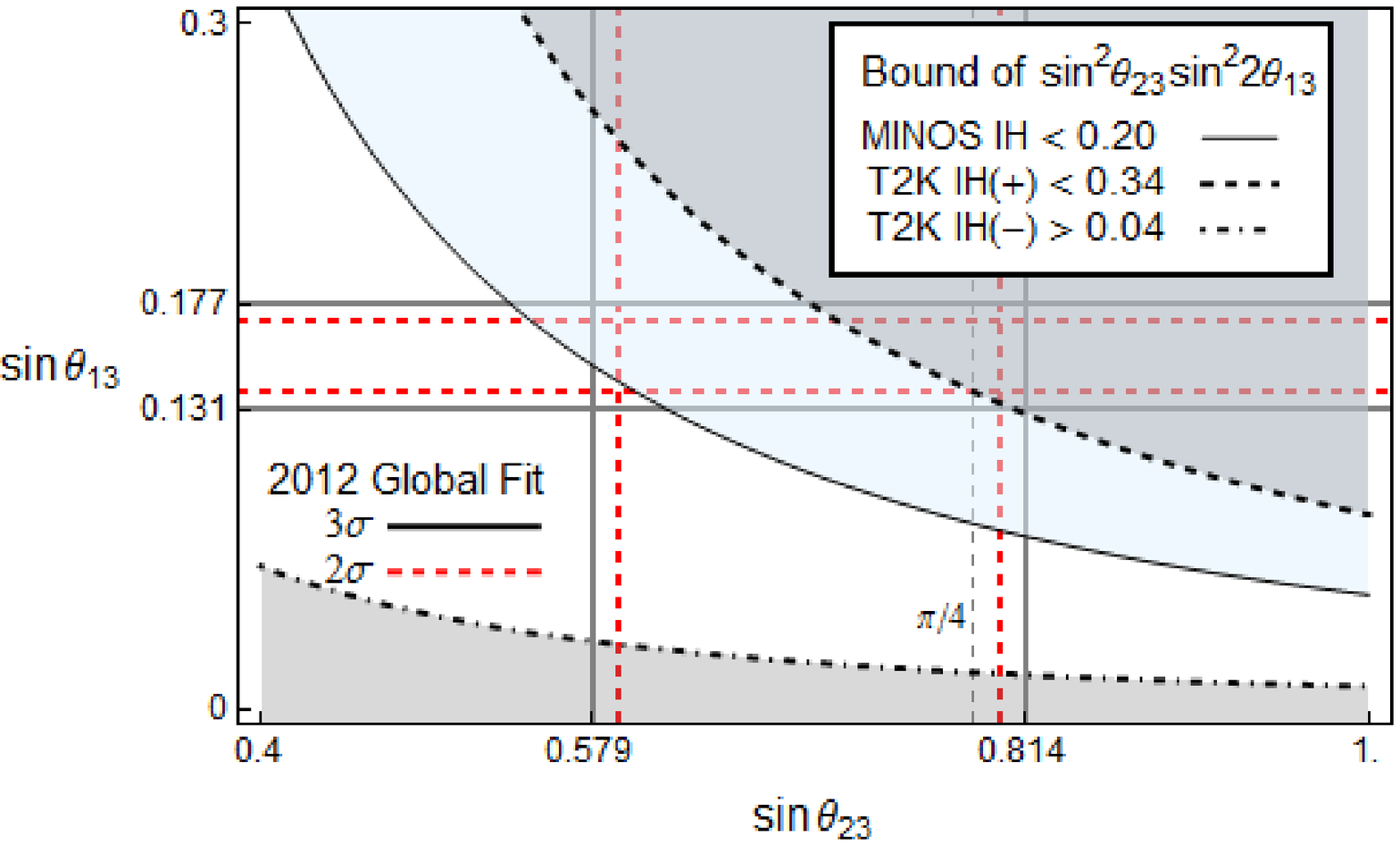}}
\caption{\label{fig1:exclusion}
Bounds of $\sin^2\theta_{23}\sin^22\theta_{13}$ as results of T2K and MINOS. Cases of normal hierarchy and inverted hierarchy are drawn separately. The $2\sigma-$ and $3\sigma-$ ranges of $\sin\theta_{13}$ and $\sin\theta_{23}$ are given in a $3\nu$ global analysis obtained in 2012 by Fogli et al \cite{Fogli:2012ua}.}
\end{figure}

Most recent announcements from long-baseline oscillations T2K\cite{Abe:2011sj} and MINOS\cite{Adamson:2011qu} and include the measurements of $2\sin^2\theta_{23}\sin^2 2\theta_{13}$, which is the leading term in the appearance probability $P(\nu_\mu\rightarrow\nu_e)$. The results from MINOS provided its bound for normal(inverted) mass hierarchy such as $2\sin^2\theta_{23}\sin^2 2\theta_{13} < 0.12~(0.20)$ at 90\% confidence level(CL) for $\delta=0$. The subsequent announcement from T2K contains $0.03(0.04) < 2\sin^2\theta_{23}\sin^2 2\theta_{13} < 0.28(0.34)$ at 90\% CL for the normal(inverted) hierarchy and $\delta=0$. In Fig.\ref{fig1:exclusion}, the bound of $2\sin^2\theta_{23}\sin^2 2\theta_{13}$ is described for each type of mass hierarchy. The areas excluded by experiments are shaded with solid boundaries for MINOS and dashed boundaries for T2K. The MINOS upper bound excludes most of the allowed space in $\sin\theta_{23}-\sin\theta_{13}$, although the estimation of the allowed regions of $\sin\theta_{23}$ and $\sin\theta_{13}$ includes the results of MINOS and T2K themselves. The $3\nu$ analysis presents the correlated analysis of $\sin^2\theta_{23}$ and $\sin^2\theta_{13}$ while two ranges of the angles in Fig.\ref{fig1:exclusion} are given independently. As shown in the figure, the vanishing $U_{e3}$ has been ruled out, and the angle $\theta_{23}$ in the first octant is strongly preferred.

As for the mass hierarchy, the NH in the figure has a disadvantage within the allowed ranges of parameter space. In the NH case in Fig.\ref{fig1:exclusion}, the allowed ranges of two angles are ruled out by the MINOS exclusion curve. Even in the IH case of Fig.\ref{fig1:exclusion}, only the small corner of the allowed box survives from the MINOS exclusion. Furthermore, SK and MINOS reported the preference of IH by using a strategy comparing the standard deviations such as $\chi^2_{IH}<\chi^2_{NH}$ \cite{Lee:2012Kyoto,Qiu:2012Kyoto}. The degeneracy problems, sign$(\theta_{23}-\pi/4)$, sign$(m_3^2-m_2^2)$, and $\sin\theta_{13}\cos\delta$, are taking directions to the solutions, including the rough indication for $\delta=0.89\pi(0.90\pi)$ within $1\sigma$-range, $0.45(0.47)-1.18(1.22)$ as presented in Ref.\cite{Fogli:2012ua}. In the rest of this section, we introduce an ansatz for $\theta_{23}$ in the PMNS to stay in the first octant.

The PMNS matrix is given as $U_{PMNS} = U_l^\dagger U_\nu$, where $U_l$ is the transformation of left-handed charged leptons and $U_\nu$ is the transformation of massive neutrinos from weak interaction basis to mass basis. In the standard parametrization, the elements of $U_{PMNS}$ are expressed in terms of physical mixing angles and a phase
\begin{widetext}
\begin{eqnarray}
	U_{PMNS}=\left(\begin{matrix}	
		c_{12}c_{13} & c_{13}s_{12} & s_{13}e^{-i\delta} \\
		-c_{23}s_{12}-c_{12}s_{13}s_{23}e^{i\delta} &
		c_{12}c_{23}-s_{12}s_{13}s_{23}e^{i\delta} &
		c_{13} s_{23} \\
		s_{23}s_{12}-c_{12} c_{23} s_{13}e^{i\delta} &
		-c_{12} s_{23} -c_{23}s_{12} s_{13}e^{i\delta} &
		c_{13} c_{23}
	\end{matrix}\right),\label{standard}
\end{eqnarray}
\end{widetext}
where $c_{ij}$ and $s_{ij}$ are $\cos\theta_{ij}$ and $\sin\theta_{ij}$, respectively, and $\delta_{CP}$ is the Dirac phase. The unitary transformation of the three-generation neutrinos is given by
    \begin{eqnarray}
    \widetilde{U}_\nu &=& R(\theta_{23})R(\theta_{13},\delta_\nu)R(\theta_{12})P(\phi, \phi') \\
    &\equiv& U_\nu P(\phi, \phi'),
    \end{eqnarray}
where the $R$'s operations are the rotations with three mixing angles and a Dirac phase $\delta_\nu$, while $P=Diag(e^{i\phi}, e^{i\phi'}, 1)$ with Majorana phases $\phi$ and $\phi'$ is a diagonal phase transformation.

The neutrino mixing $U_\nu$ is chosen as a bi-maximal, $U_\nu=U_{BM}$, where
    \begin{eqnarray}
        U_{BM} &=& \left(
        \begin{array}{ccc}
        \frac{1}{\sqrt{2}} & \frac{1}{\sqrt{2}} & 0 \\
        -\frac{1}{2} & \frac{1}{2} & \frac{1}{\sqrt{2}} \\
        \frac{1}{2} & -\frac{1}{2} & \frac{1}{\sqrt{2}}
        \end{array} \right),\label{bimaximal}
    \end{eqnarray}
with $\sin\theta_{12}=1/\sqrt{3}, \sin\theta_{23}=1/\sqrt{2}$, and $\sin\theta_{13}=0$. If the mixing of charged leptons $U_l$ is a single-angle rotation of the 1-2 block of the mass matrix such that
    \begin{eqnarray}
        U_l = \left(
        \begin{array}{ccc}
        \cos\chi & \sin\chi e^{-i\varphi} & 0 \\
        -\sin\chi e^{i\varphi} & \cos\chi & 0 \\
        0 & 0 & 1
        \end{array} \right),\label{rotation12}
    \end{eqnarray}
then, the PMNS matrix obtained by $U_l^\dagger U_{BM}$ is given as
    \begin{eqnarray}
        U_A =
        \left(
        \begin{array}{ccc}
        \frac{\cos\chi}{\sqrt{2}}+\frac{\sin\chi}{2}e^{-i\varphi} &
        \frac{\cos\chi}{\sqrt{2}}-\frac{\sin\chi}{2}e^{-i\varphi} &
        -\frac{\sin\chi}{\sqrt{2}}e^{-i\varphi} \\
        -\frac{\cos\chi}{2}+\frac{\sin\chi}{\sqrt{2}}e^{i\varphi} &
        \frac{\cos\chi}{2}+\frac{\sin\chi}{\sqrt{2}}e^{i\varphi} &
        \frac{\cos\chi}{\sqrt{2}} \\
        \frac{1}{2} & -\frac{1}{2} & \frac{1}{\sqrt{2}}
        \end{array} \right).\label{pmns12}
    \end{eqnarray}
When the elements of $U_A$ are compared with $U_{PMNS}$ in Eq. {\ref{standard}}, the three angles in the PMNS matrix are described by the following simple relations
    \begin{eqnarray}
    && s_{13} = \frac{1}{\sqrt{2}}\sin\chi \label{case1chi}\\
    && s_{23} = \frac{\cos\chi}{\sqrt{2-\sin^2\chi}} \nonumber \\
    && s_{12} = \frac{\cos\chi - \sin\chi/\sqrt{2}}{\sqrt{2-\sin^2\chi}},
    \nonumber
    \end{eqnarray}
where $\delta=\pi$ in Eq. (\ref{standard}) and $\varphi=0$ in Eq. (\ref{rotation12}). Eliminating $\chi$, the $s_{23}$ and $s_{12}$ can be expressed in terms of a single variable $s_{13}$ and $1/\sqrt{2}$ as
    \begin{eqnarray}
    s_{23} &=& \sqrt{\frac{1-2s_{13}^2}{2-2s_{13}^2}} \label{case1theta23} \\
    s_{12} &=& \frac{\sqrt{1-2s_{13}^2}-s_{13}}{\sqrt{2-2s_{13}^2}},
    \label{case1theta12}
    \end{eqnarray}
or the three angles constrain themselves by the relation,
    \begin{eqnarray}
    s_{23}=s_{12}+\frac{1}{\sqrt{2}}\tan\theta_{13}.
    \end{eqnarray}
Since $s_{23}$ and $s_{12}$ decrease from $1/\sqrt{2}$ as $s_{13}$ increases from 0, the angle $\theta_{23}$ that belongs to the first octant is a consequence of the ansatz in Eq.(\ref{pmns12}).

\begin{figure}
\resizebox{90mm}{!}{\includegraphics[width=0.75\textwidth]{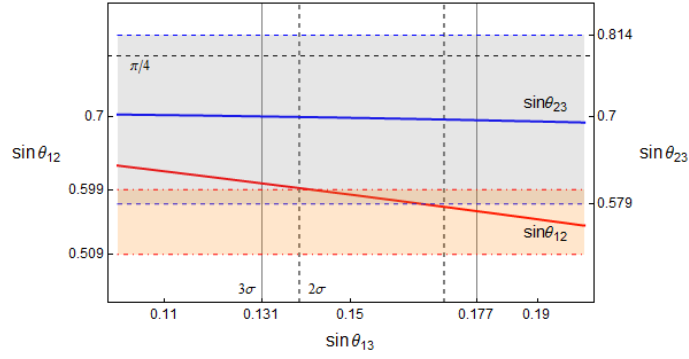}}
\caption{\label{fig2:ansatz} (blue) $\sin\theta_{23}$ given in Eq.(\ref{case1theta23}) and (red) $\sin\theta_{12}$ given in Eq.(\ref{case1theta12}). The allowed ranges at $3\sigma$ CL in $\sin\theta_{23}$ and $\sin\theta_{12}$ are described by the gray and orange shades, respectively.}
\end{figure}

For comparison, it is worth considering the case in which $U_l$ is a another single-angle rotation of the 1-3 block of the mass matrix such that
    \begin{eqnarray}
        U_l = \left(
        \begin{array}{ccc}
        \cos\chi & 0 & \sin\chi e^{-i\varphi} \\
        0 & 1 & 0 \\
        -\sin\chi e^{i\varphi} & 0 & \cos\chi
        \end{array} \right).\label{rotation13}
    \end{eqnarray}
Then the PMNS matrix which is obtained by $U_l^\dagger U_{BM}$ is:
    \begin{eqnarray}
        U_B =
        \left(
        \begin{array}{ccc}
        \frac{\cos\chi}{\sqrt{2}}-\frac{\sin\chi}{2}e^{-i\varphi} &
        \frac{\cos\chi}{\sqrt{2}}+\frac{\sin\chi}{2}e^{-i\varphi} &
        -\frac{\sin\chi}{\sqrt{2}}e^{-i\varphi} \\
        -\frac{1}{2} & \frac{1}{2} & \frac{1}{\sqrt{2}} \\
        \frac{\cos\chi}{2}+\frac{\sin\chi}{\sqrt{2}}e^{i\varphi} &
        -\frac{\cos\chi}{2}+\frac{\sin\chi}{\sqrt{2}}e^{i\varphi} &
        \frac{\cos\chi}{\sqrt{2}}
        \end{array} \right).\label{pmns13}
    \end{eqnarray}
When the elements of $U_B$ are compared with $U_{PMNS}$ in Eq. {\ref{standard}}, the three angles in the PMNS matrix are described by the following simple relations
    \begin{eqnarray}
    && s_{13} = \frac{1}{\sqrt{2}}\sin\chi \label{case2chi}\\
    && s_{23} = \frac{1}{\sqrt{2-\sin^2\chi}} \nonumber \\
    && s_{12} = \frac{\cos\chi -\sin\chi/\sqrt{2}}{\sqrt{2-\sin^2\chi}},
    \nonumber
    \end{eqnarray}
where $\delta=0$ in Eq. (\ref{standard}) and $\varphi=\pi$ in Eq. (\ref{rotation13}). Eliminating $\chi$, the $\sin\theta_{23}$ in terms of $\theta_{13}$ is obtained as
    \begin{eqnarray}
    s_{23} &=& \frac{1}{\sqrt{2-2s_{13}^2}} \label{case2theta23}.
    \label{case2theta12}
    \end{eqnarray}
In the above case, the angle $\theta_{23}$ belongs to the second octant, as a consequence of the ansatz in Eq.(\ref{pmns13}).

\section{Model for the $\theta_{23}$ in the first octant and the inverted mass hierarchy}

We introduce a model that can derive the ansatz in Eq.(\ref{pmns12}) within a simple framework in the sense that the discrete flavor symmetry is $\mathbb{S}_3$. The field contents and their representations under $\mathbb{S}_3\otimes\mathbb{Z}_2$ are listed in Table \ref{sm_rep} and Table \ref{beyondsm_rep}. The $\mathbb{Z}_2$ symmetry plays a role in keeping the particles in the SM from coupling with additional Higgs beyond the SM. As shown in Table \ref{sm_rep} and Table \ref{beyondsm_rep}, all the SM particles and the Higgs to couple with them are $\mathbb{Z}_2$ even, while the right-handed neutrinos and the Higgs to couple with them are all $\mathbb{Z}_2$ odd.

    \begin{table}
    \caption{Group representations of fermions. The SU(2) representation and hypercharge of a field are denoted by the subscription `$G$' of the gauge symmetry, and the $\mathbb{S}_3$ representation and  $\mathbb{Z}_2$ charge of the field are denoted by the subscription `$F$'.}
    \begin{ruledtabular}
    \begin{tabular}{c|ccc|ccc}
        $\mathrm{Rep.}$  & $(\mathbf{1},1)_F$ & $(\mathbf{2},1)_F$ &  & $(\mathbf{1'},-1)_F$ &  & $(\mathbf{1},-1)_F$  \\ \hline
        $(\mathbf{2},-1/2)_G$  & $l_e$ & $L_\alpha : (l_\mu, l_\tau)$ & & ... & & ... \\
        $(\mathbf{1},-1)_G$  & $e_r$ & $R_\alpha : (\mu_r, \tau_r)$ & & ... & & ...\\
        $(\mathbf{1},0)_G$  & ... & ... & & $n_1, ~n_2$ &  & $n_3$ \\
    \end{tabular} \label{sm_rep}
    \end{ruledtabular}
    \caption{Group representations of Higgs scalars. An Abelian discrete symmetry $\mathbb{Z}_2$ is also adopted to distinguish extra particles from the SM contents. The $\mathbb{Z}_2$-odd Higgs do not interact with $\mathbb{Z}_2$-even SM fields at tree level.}
    \begin{ruledtabular}
    \begin{tabular}{c|ccc|ccc}
        $\mathrm{Rep.}$  & $(\mathbf{1},1)_F$ & $(\mathbf{1},1)_F$ &  & $(\mathbf{1'},-1)_F$ & & $(\mathbf{2},-1)_F$  \\ \hline
        $(\mathbf{2},-1/2)_G$  & $H$ & $\Phi : (\phi_1, \phi_2)$ & & $h$ & & $\Sigma : (\sigma_1, \sigma_2)$ \\
    \end{tabular} \label{beyondsm_rep}
    \end{ruledtabular}
    \end{table}

\subsection{Charged lepton masses with a single mixing $\theta_{12}$ }

The Lagrangian of Yukawa couplings of the charged leptons and the Higgs scalar doublet $H$ and $\Phi$ is
    \begin{eqnarray}
    -\mathcal{L}_{SM} = c_1 H \overline{e}_r l_e + c_2 H \overline{R}_\alpha L_\alpha
    + c_3 \Phi \overline{R}_\alpha L_\alpha + c_4 \Phi \overline{e}_r L_\alpha + c_5 \Phi \overline{R}_\alpha l_e, \label{yukawa}
    \end{eqnarray}
where the SU(2) fermion doublet $l_e$ is a flavor singlet but $l_\mu$ and $l_\tau $ belong to a doublet $L_\alpha\equiv(l_\mu, l_\tau)$ under $\mathbb{S}_3$. Also, the righthanded charged lepton singlet $e_r$ is a flavor singlet while $\mu_r$ and $\tau_r$ belong to a doublet $R_\alpha\equiv(\mu_r, \tau_r)$. The Higgs scalar doublet $H$ of SM is involved in the above interactions as a flavor singlet, while the additional Higgs scalar doublet $\Phi$ is a flavor doublet with component fields $(\varphi_1, \varphi_2)$. The Dirac mass matrix of the charged leptons from the Yukawa couplings becomes
    \begin{eqnarray}
        M_A^l \sim \left(
        \begin{array}{ccc}
        c_1 v & c_4 v_1 & 0 \\
        c_5 v_1 & c_2 v - c_3 v_1 & 0 \\
        0 & 0 & c_2 v + c_3 v_1
        \end{array} \right),\label{mass_lepton_12mix}
    \end{eqnarray}
where the vev $\langle\Phi\rangle$ is $(v_1, 0)$, as explained in the Appendix.

The masses of leptons are obtained from $U_l M_l^\dagger M_l U_l^\dagger = Diag(m_e^2, m_\mu^2, m_\tau^2)$. We denote the matrix $M_l^\dagger M_l$ by $K$ as
\begin{widetext}
    \begin{eqnarray}
    K \equiv \left(
        \begin{array}{ccc}
        |c_1|^2 v^2 + |c_5|^2 v_1^2 & c_1^*c_4 v v_1 + c_5^* v_1 (c_2 v - c_3 v_1) & 0 \\
        c_v^*c_1 v v_1 + (c_2^* v - c_3^* v_1)c_5 v_1  & |c_2 v - c_3 v_1|^2 & 0 \\
        0 & 0 & |c_2 v + c_3 v_1|^2
        \end{array} \right), \label{mdaggerm}
    \end{eqnarray}
\end{widetext}
which is plausible by the relation in terms of a single mixing angle, a phase, and masses as in $K=R(\theta_l, \delta_l)Diag(m_e^2, m_\mu^2, m_\tau^2)R^\dagger(\theta_l, \delta_l)$, where the 1-2 block rotation $R(\theta_l, \delta_l)$ is given by
    \begin{eqnarray}
    R(\theta_l, \delta_l) \equiv \left(
        \begin{array}{ccc}
        \cos\theta_l & \sin\theta_l e^{-i\delta_l} & 0 \\
        -\sin\theta_l e^{i\delta_l} & \cos\theta_l & 0 \\
        0 & 0 & 1
        \end{array} \right).\label{TansLepton}
    \end{eqnarray}
The elements of the matrix $K$ in Eq.(\ref{mdaggerm}) are described by physical parameters,
    \begin{eqnarray}
    K_{11} &=& m_e^2\cos^2\theta_l + m_\mu^2\sin^2\theta_l, \nonumber \\
    K_{22} &=& m_\mu^2\cos^2\theta_l + m_e^2\sin^2\theta_l, \label{elementK}\\
    K_{12} &=& K_{21}^* = \left( m_\mu^2e^{i\delta_l} - m_e^2e^{-i\delta_l}\right)\cos\theta_l\sin\theta_l \nonumber \\
    K_{33} &=& m_\tau^2. \nonumber
    \end{eqnarray}
In the opposite way, the mixing angle $\theta_l$ and the phase $\delta_l$ are obtained from the elements in Eq.(\ref{elementK}) as
    \begin{eqnarray}
    \tan2\theta_l \cos\delta_l= \frac{K_{12}+K_{12}^*}{K_{22}-K_{11}},\label{thetadelta}
    \end{eqnarray}
or from the matrix in Eq.(\ref{mdaggerm}):
    \begin{eqnarray}
    \tan2\theta_l = \frac{2 \mathrm{Re}[c_1^*c_4 v v_1 + c_5^* v_1 (c_2 v - c_3 v_1)]}{|c_2 v - c_3 v_1|^2-|c_1|^2 v^2 - |c_5|^2 v_1^2}.\label{theta}
    \end{eqnarray}
In general, the squared masses can be expressed in the following way:
    \begin{eqnarray}
    m_e^2 &=& \frac{1}{2}\left( K_{22} + K_{11} \right)
            - \frac{1}{2}\left( K_{22} - K_{11} \right)\sqrt{1 +\tan^2 2\theta_l\cos^2\delta_l}, \nonumber \\
    m_\mu^2 &=& \frac{1}{2}\left( K_{22} + K_{11} \right)
            + \frac{1}{2}\left( K_{22} - K_{11} \right)\sqrt{1 +\tan^2 2\theta_l\cos^2\delta_l},
    \label{msquare}
    \end{eqnarray}
and $m_\tau^2 = |c_2 v + c_3 v_1|^2$.

\subsection{Light neutrino masses with bi-maximal mixing}

The Majorana masses of the gauge singlets $n_i$, in $\frac{1}{2} M_1 n_1 n_1 + \frac{1}{2} M_2 n_2 n_2 + \frac{1}{2} M_3 n_3 n_3 + M_\mathrm{x} n_1 n_2$, are expressed in the following matrix:
    \begin{eqnarray}
    M_R =  \left(
        \begin{array}{ccc}
        M_{11} & M_\mathrm{x} & 0 \\
        M_\mathrm{x} & M_{22} & 0 \\
        0 & 0 & M_3
        \end{array} \right), \label{majorana}
    \end{eqnarray}
where the flavor charges are listed in Table \ref{sm_rep}.
All additional fields beyond the SM, including righthanded neutrinos, are distinguished from the SM particles by $\mathbb{Z}_2$ parity. All the SM fields are $\mathbb{Z}_2$ even, so the parity does not affect any interaction of SM particles. Additional Higgs scalars, $\Sigma \equiv (\sigma_1,\sigma_2)$ and $h$, all have $\mathbb{Z}_2$-odd quantum number. Their representations under the gauge symmetry and those under the flavor symmetry are listed in Table \ref{sm_rep} and Table \ref{beyondsm_rep}. The Yukawa couplings of the right-handed neutrinos are given as
    \begin{eqnarray}
    -\mathcal{L}_\mathrm{neutrino} = f_0 h \overline{n}_1 l_e + f_0' h \overline{n}_2 l_e
    + f_1 \Sigma \overline{n}_1 L_\alpha + f_2 \Sigma \overline{n}_2 L_\alpha + f_3 \Sigma \overline{n}_3 L_\alpha. \label{yukawa_neutrino}
    \end{eqnarray}
If the product of the $\mathbb{S}_3$ doublet $\Sigma$ with  $L_\alpha$ is the representation $\mathbf{1}$, it can be decomposed into $\sigma_1 l_\mu + \sigma_2 l_\tau$, or if the product is the representation $\mathbf{1'}$, it can be decomposed into $\sigma_1 l_\tau - \sigma_2 l_\mu$. Their Yukawa interactions can be expressed by the following Yukawa matrix:
    \begin{eqnarray}
        Y =  \left(
        \begin{array}{ccc}
        f_0h & f_0'h & 0 \\
        -f_1\sigma_2 & -f_2\sigma_2 & f_3\sigma_1 \\
        f_1\sigma_1 & f_2\sigma_1 & f_3\sigma_2
        \end{array} \right). \label{yuk_ai}
    \end{eqnarray}

The seesaw mechanism with the Majorana mass in Eq.(\ref{majorana}) and the Yukawa matrix $Y$ in Eq.(\ref{yuk_ai}) results in
the mass matrix of light neutrinos as follows:
\begin{widetext}
    \begin{eqnarray}
        M_\nu = \mathrm{S_0} \left(
            \begin{array}{ccc}
            1 & 0 & 0 \\
            0 & 0 & 0 \\
            0 & 0 & 0
            \end{array} \right)
        + \mathrm{S_1} \left(
            \begin{array}{ccc}
            0 & 1 & -1 \\
            1 & 0 & 0 \\
            -1 & 0 & 0
            \end{array} \right)
        + \mathrm{S_2} \left(
            \begin{array}{ccc}
            0 & 0 & 0 \\
            0 & 1 & -1 \\
            0 & -1 & 1
            \end{array} \right)
        + \mathrm{S_2'} \left(
            \begin{array}{ccc}
            0 & 0 & 0 \\
            0 & 1 & 1 \\
            0 & 1 & 1
            \end{array} \right), \label{seesaw}
    \end{eqnarray}
\end{widetext}
where
    \begin{eqnarray}
    S_0 &=& \frac{f_0^2u^2}{\mu_1} - \frac{2f_0f_0'u^2}{\mu_{\mathrm{x}}} + \frac{f_0'^2u^2}{\mu_2} \label{seesaw_ih} \\
    S_1 &=& \frac{f_0f_1uw}{\mu_1} - \frac{(f_0f_2+f_0'f_1)uw}{\mu_{\mathrm{x}}} + \frac{f_0'f_2uw}{\mu_2} \nonumber \\
    S_2 &=& \frac{f_1^2w^2}{\mu_1} - \frac{2f_1f_2w^2}{\mu_{\mathrm{x}}} + \frac{f_2^2w^2}{\mu_2} \nonumber \\
    S_2' &=& \frac{f_3^2w^2}{M_3}. \nonumber
    \end{eqnarray}
The Majorana mass of the right-handed neutrinos contribute to the lightness of neutrinos as
    \begin{eqnarray}
    \mu_1^{-1} &\equiv& M_{22}\left( M_{11}M_{22} -M_{\mathrm{x}}^2 \right)^{-1} \\
    \mu_2^{-1} &\equiv& M_{11}\left( M_{11}M_{22} -M_{\mathrm{x}}^2 \right)^{-1} \nonumber \\
    \mu_\mathrm{x}^{-1} &\equiv& M_{\mathrm{x}}\left( M_{11}M_{22} -M_{\mathrm{x}}^2 \right)^{-1}. \nonumber
    \end{eqnarray}

If the transformation $U_\nu$ is exactly bi-maximal, the symmetric mass matrix of light neutrinos obtained by $M_\nu=U_{BM}Diag(m_1, m_2, m_3)U_{BM}^T$ has the form in Eq.(\ref{seesaw}) with $S_0=(\check{m}_1 + \check{m}_2)/2,~S_1=(\check{m}_2 - \check{m}_1)/2\sqrt{2},~S_2=(\check{m}_1 + \check{m}_2)/4$, and $S_2'=m_3/2$. In comparison with $S_0, S_1, S_2$ and $S_2'$ in Eq.(\ref{seesaw_ih}), the following relations are obtained:
    \begin{eqnarray}
    \check{m}_2 + \check{m}_1 &=& 2 \left( \frac{f_0^2u^2}{\mu_1} - \frac{2f_0f_0'u^2}{\mu_{\mathrm{x}}} + \frac{f_0'^2u^2}{\mu_2} \right) ~\mathrm{or} \nonumber \\
        &=& 2 \left( \frac{f_1^2w^2}{\mu_1} - \frac{2f_1f_2w^2}{\mu_{\mathrm{x}}} + \frac{f_2^2w^2}{\mu_2} \right) \\
    \check{m}_2 - \check{m}_1 &=& 2\sqrt{2} \left( \frac{f_0f_1uw}{\mu_1} - \frac{(f_0f_2+f_0'f_1)uw}{\mu_{\mathrm{x}}} + \frac{f_0'f_2uw}{\mu_2} \right)  \\
    m_3 &=& 2 \frac{f_3^2w^2}{M_3},
    \end{eqnarray}
which provide the inverted order of mass hierarchy if $M_3 \gg M_1,~M_2$.

\section{Conclusion}
Recent updates in the results of neutrino oscillations give rise to the preference for the $\theta_{23} < \pi/4$, the sizable $\theta_{13}$, the inverted mass hierarchy, and the CP phase $\delta=\pi$. The information will dismantle the degeneracies among neutrino parameters in the near future.

We propose an ansatz motivated by the significant experimental advance, in which the PMNS matrix is obtained by the bi-maximal neutrino mixing and a single rotation of the 1-2 block of charged lepton masses. All the mixing angles $\sin\theta_{ij}$ in the PMNS matrix are expressed in terms of the single angle of $U_l$ and $1/\sqrt{2}$, so that there can be a relation established among the three angles in the PMNS as follows:
    \begin{eqnarray}
    \tan\theta_{13}=\sqrt{2}\left(\sin\theta_{23}-\sin\theta_{12}\right).
    \end{eqnarray}
The $\theta_{23}$ in the first octant is obtained specifically when the $U_l$ is the single rotation of the 1-2 block. If the $U_l$ is given by the single rotation of the 1-3 block, the $\theta_{23}$ cannot be smaller than $\pi/4$, becoming larger as $\theta_{13}$ increases.  That is, the octant of $\theta_{23}$ is determined as a consequence of the ansatz.

We proposed a model that can produce the ansatz. The flavor symmetry $\mathbb{S}_3\otimes\mathbb{Z}_2$ has been adopted. The mass matrix of charged leptons are diagonalized by a single angle, while the matrix of the light neutrino masses are bi-maximally mixed. The mass hierarchy is characterized by the inverted order with the smallest $m_3$.

\begin{acknowledgments}
This work was supported by the Basic Science Research program through the NRF(2012-0004311).
\end{acknowledgments}

\appendix

\section{Discrete flavor symmetry $\mathbb{S}_3$}

The Lagrangians in Eq.(\ref{yukawa}) and Eq.(\ref{yukawa_neutrino}) are constructed under the minimal non-Abelian discrete symmetry $\mathbb{S}_3$. There are six elements of the group in three classes, and their irreducible representations are $\mathbf{1},~ \mathbf{1'},$ and $\mathbf{2}.$ Its character table is mentioned in many models \cite{Morisi:2011pm,Meloni:2010aw,Dong:2011vb,Chu:2011jg}.

The Clebsch-Gordon coefficients in the real representations are given by the following product rules \cite{Ma:2004pt,Chen:2004rr},
    \begin{eqnarray}
    && \mathbf{1'}\times\mathbf{1'} = \mathbf{1}~:~ ab, \label{prod1}\\
    && \mathbf{1'}\times\mathbf{2} = \mathbf{2}~:~
        \left(\begin{array}{r}ab_2\\-ab_1 \end{array}\right)\\
    && \mathbf{2}\times\mathbf{2} =
        \mathbf{1}+\mathbf{1'}+\mathbf{2}, \label{prod3} \\
    && \begin{array}{lcl}
        \mathbf{1} & : & (a_1b_1+a_2b_2) \\
        \mathbf{1'} & : & (a_1b_2-a_2b_1) \\
        \mathbf{2} & : &
        \left(\begin{array}{l}a_2b_2-a_1b_1\\a_1b_2+a_2b_1 \end{array}\right). \nonumber
        \end{array} \label{prod5}
    \end{eqnarray}

\section{Higgs Potential}\footnote{The Higgs contents and their self potential are identical with those in previous work \cite{Siyeon:2012zu}. Most parts of details have been skipped to avoid the repetition.}

The contents of Higgs scalar particles and their representations under $\mathbb{S}_3\otimes\mathbb{Z}_2$ are
    \begin{eqnarray}\begin{array}{lll}
        (\mathbf{1},1)_F & : & H \\
        (\mathbf{2},1)_F & : & \Phi~(\varphi_1, \varphi_2) \\
        (\mathbf{1},-1)_F & : & h \\
        (\mathbf{2},-1)_F & : & \Sigma~(\sigma_1, \sigma_2), \label{4higgs}
    \end{array}
    \end{eqnarray}
which commonly belong to $(\mathbf{2},1/2)_G$ under the $SU(2)\times U(1)$ gauge group.
The full invariant Higgs potential can be organized into three parts as follows:
    \begin{eqnarray}
    V=V_e(H,\Phi)+V_o(h,\Sigma)+V_\chi(H,\Phi;h,\Sigma), \label{3potential}
    \end{eqnarray}
where $V_e$ and $V_o$ are the interactions of only $\mathbb{Z}_2$-even particles and those of only $\mathbb{Z}_2$-odd particles, respectively, while $V_\chi$ is the cross interactions of $\mathbb{Z}_2$-even and $\mathbb{Z}_2$-odd particles. Each contribution to the potential $V$ is given as:
\begin{widetext}
    \begin{eqnarray}
        && V_e(H,\Phi) = \label{potentialNH}
            m_H^2H^\dagger H + \frac{1}{2}\eta(H^\dagger H)^2 + m_\varphi^2 \Phi^\dagger\Phi +\frac{1}{2}\Lambda(\Phi^\dagger\Phi)_r^2 \\
        && \hspace{10pt}+ ~\lambda(\Phi^\dagger\Phi)_1(H^\dagger H)_1 +
            \lambda'(\Phi^\dagger H)_2(H^\dagger\Phi)_2+\lambda''\{(\Phi^\dagger H)_2^2 + h.c.\} +\kappa\{(\Phi^\dagger\Phi)_2(\Phi^\dagger H)_2 + h.c.\}, \nonumber \\
            \nonumber \\
        && V_o(h,\Sigma)  =  \label{potential_sigma}
            m_h^2h^\dagger h + \frac{1}{2}\lambda_h(h^\dagger h)^2
            + m_s^2 \Sigma^\dagger\Sigma +
            \frac{1}{2}\Lambda_s(\Sigma^\dagger\Sigma)_r^2 \\
        && \hspace{10pt}+ ~\lambda_s(\Sigma^\dagger\Sigma)_1(h^\dagger h)_1 +
            \lambda_s'(\Sigma^\dagger h)_2(h^\dagger\Sigma)_2 +
            \lambda_s''\{(\Sigma^\dagger h)_2^2+\mathrm{h.c.}\} + \kappa_s\{(\Sigma^\dagger\Sigma)_2(\Sigma^\dagger h)_2+\mathrm{h.c.}\}.\nonumber \\
            \nonumber \\
        && V_\chi(H,\Phi;h,\Sigma) =  \label{potential_ext}
        \chi(H^\dagger H)_1(h^\dagger h)_1+\chi'(H^\dagger h)_1(h^\dagger H)_1+
            \chi''\{(H^\dagger h)_1^2+\mathrm{h.c.}\} \\
        && \hspace{10pt}+ ~\lambda_\chi(\Phi^\dagger \Phi)_1(h^\dagger h)_1 +
            \lambda_\chi'(\Phi^\dagger h)_2(h^\dagger \Phi)_2+
            \lambda_\chi''\{(\Phi^\dagger h)_2^2+\mathrm{h.c.}\} \nonumber \\
        && \hspace{10pt}+ ~\eta_\chi(\Sigma^\dagger\Sigma)_1(H^\dagger H)_1 +
            \eta_\chi'(\Sigma^\dagger H)_2(H^\dagger \Sigma)_2+
            \eta_\chi''\{(\Sigma^\dagger H)_2^2+\mathrm{h.c.}\} \nonumber \\
        && \hspace{10pt}+ ~\gamma\{(H^\dagger h)_1(\Sigma^\dagger\Phi)_1+\mathrm{h.c.}\} +
            \gamma'\{(H^\dagger \Sigma)_2(h^\dagger\Phi)_2+\mathrm{h.c.}\} \nonumber \\
        && \hspace{10pt}+ ~\Gamma_\chi(\Phi^\dagger \Phi)_r(\Sigma^\dagger\Sigma)_r +
            \Gamma_\chi'(\Phi^\dagger\Sigma)_r(\Sigma^\dagger\Phi)_r+
            \Gamma_\chi''\{(\Phi^\dagger \Sigma)_r^2+\mathrm{h.c.}\}.
            \nonumber
    \end{eqnarray}
\end{widetext}
The subscripts `1' and `2' in each term indicate that the product of the two fields belongs to the representation $\mathbf{1}$ or $\mathbf{2}$ in $\mathbb{S}_3$. Each term with a subscript `$r$' consists of three types of products, $\mathbf{1}, ~\mathbf{1'}$ and $\mathbf{2}$ representations. According to the product rules in Eqs. (\ref{prod1}) - (\ref{prod3}),
$(\Phi^\dagger\Phi)_1=|\varphi_1|^2+|\varphi_2|^2,~(\Phi^\dagger\Phi)_{1'}=\varphi_1^*\varphi_2-\varphi_2^*\varphi_1$,
and $(\Phi^\dagger\Phi)_2=(|\varphi_2|^2-|\varphi_1|^2 ~ ~ \varphi_1^*\varphi_2+\varphi_2^*\varphi_1)^T$.
The  Higgs potential in Eq.(\ref{potentialNH}) can be rephrased in terms of component fields $\{\varphi_i, \varphi_i^\dagger\}$ with $i=1$ and 2, and $\{H, H^\dagger\}$.
When the Higgs particles obtain their real vacuum expectation values such that $\langle H\rangle=\langle H^\dagger\rangle=v$,  $\langle\varphi_1\rangle=v_1$, and $\langle\varphi_2\rangle=v_2$, the potential can be expressed as follows.
    \begin{eqnarray}
    && V_e~(v,v_1,v_2) ~= ~m_H^2v^2 + m_\varphi^2(v_1^2+v_2^2) + \frac{1}{2}\eta v^4 \label{potential_vev} \\
    && \hspace{10pt}+ ~\frac{1}{2}\Lambda_a(v_1^2+v_2^2)^2 +\Lambda_b v^2(v_1^2+v_2^2)
    + 2\kappa v(3v_2^2v_1-v_1^3), \nonumber
    \end{eqnarray}
where $\Lambda_a=\lambda_a+\lambda_c$, and $\Lambda_b=\lambda+\lambda'+2\lambda''$.
Following the same steps as in Eq.(\ref{potentialNH}) - Eq.(\ref{potential_vev}), the potentials, $V_o$ and $V_\chi$, in terms of vevs, $\langle h\rangle=u$ and $(\langle \sigma_1\rangle, \langle \sigma_2 \rangle)=(w_1,w_2)$, can be expressed as follows:
\begin{widetext}
    \begin{eqnarray}
    && V_o~(u,w_1,w_2) ~= ~m_h^2u^2 + m_s^2(w_1^2+w_2^2) + \frac{1}{2}\lambda_h u^4 \label{potential_odd} \\
    && \hspace{10pt}+ ~\frac{1}{2}\Lambda_s(w_1^2+w_2^2)^2 +\Lambda_c u^2(w_1^2+w_2^2) + ~2\kappa_s u(3w_2^2w_1-w_1^3), \nonumber
    \end{eqnarray}
where $\Lambda_c \equiv \lambda_s+\lambda_s'+2\lambda_s''$.
    \begin{eqnarray}
    && V_\chi = k_1u^2v^2 + k_2u^2(v_1^2+v_2^2) +k_3v^2(w_1^2+w_2^2) +k_4 uv(v_1w_1+v_2w_2)
            \label{potential_cross} \\
    && \hspace{10pt}+ k_5v_1v_2w_1w_2
        +k_5'(v_1^2+v_2^2)(w_1^2+w_2^2)+ k_5''(v_2^2-v_1^2)(w_2^2-w_1^2) +k_5'''(v_1^2w_1^2+v_2^2w_2^2), \nonumber
    \end{eqnarray}
\end{widetext}
where $k_1=\chi+\chi'+2\chi'', ~k_2=\lambda_\chi+\lambda_\chi'+2\lambda_\chi'', ~k_3=\eta_\chi+\eta_\chi'+2\eta_\chi'',$ and $k_4=2(\gamma+\gamma')$. The $k_5...k_5'''$ are rather complicated polynomials of $\Gamma_\chi,\Gamma'_\chi,$ and $\Gamma_\chi$ in Eq.(\ref{potential_ext}), such that $k_5=k_5(\Gamma_\chi, \Gamma'_\chi), ~k_5'=k_5'(\Gamma_\chi, \Gamma''_\chi), ~k_5''=k_5''(\Gamma_\chi),$ and $k_5'''=k_5'''(\Gamma'_\chi)$.

The first derivatives of the full potential given in Eq.(\ref{3potential}) are presented in Ref.[Siyeon]. The minimality condition is obtained by vanishing the derivatives for chosen vevs of Higgs.
The vevs, $v_1\neq0$ and $v_2=0$, can make the potential minimum, when the following conditions are necessary.
    \begin{eqnarray}
    &&\left(\frac{\partial V}{\partial v_1}\right)_{v_2=0} = 2v_1\{K(m_\varphi^2, u^2, v^2, w_i^2)+\Lambda_a v_1^2\}
        -6\kappa v v_1^2 +k_4uvw_1 =0 \nonumber \\
    &&\left(\frac{\partial V}{\partial v_2}\right)_{v_2=0} = k_4uvw_2 +k_5v_1w_1w_2 =0,
    \label{first_der}
    \end{eqnarray}
where $K(m_\varphi^2, u^2, v^2, w_i^2)$ is the part that is independent of either $v_1$ or $v_2$. The positiveness of the second derivatives are given as
    \begin{eqnarray}
    &&\left(\frac{\partial^2 V}{\partial v_1\partial v_2}\right)_{v_2=0} = k_5w_1w_2 >0
    \label{second_der} \\
    &&\left(\frac{\partial^2 V}{\partial w_1\partial w_2}\right)_{v_2=0} =
    4\Lambda_sw_1w_2 +12\kappa_s uw_2 >0. \nonumber
    \end{eqnarray}
It is clear that any of $w_1$ and $w_2$ should not be zero to satisfy the above conditions. According to the symmetry of the potential under the interchange of $\sigma_1$ and $\sigma_2$, vevs can be taken as $w_1=w_2=w$. Thus, in summary, the following vevs of the fields in Eq.(\ref{4higgs}) can be adopted for the masses of leptons:
    \begin{eqnarray}\begin{array}{lll}
        \langle H \rangle & = & v\\
        \langle \Phi \rangle & = & (v_1, 0) \\
        \langle h \rangle & = & u\\
        \langle \Sigma \rangle & = & (w, w). \label{4vevs}
    \end{array}
    \end{eqnarray}

\end{document}